\documentclass[prc,showpacs,amsmath,amssymb,manuscript]{revtex4}
\usepackage{epsfig}
\usepackage{dcolumn}
\usepackage{bm}
\usepackage{subfigure}
\input{epsf.sty}

\newcommand{\beq}{\begin{equation}}
\newcommand{\eeq}{\end{equation}}
\newcommand{\ba}{\begin{array}}
\newcommand{\ea}{\end{array}}
\newcommand{\beqa}{\begin{eqnarray}}
\newcommand{\eeqa}{\end{eqnarray}}
\newcommand{\bd}[1]{ \mbox{\boldmath $#1$}  }
\newcommand{\oh}{\frac{1}{2}}

\newcommand{\lam}{\lambda}
\newcommand{\nn}{\nonumber}

\newcommand{\macol}[2]{\left\{\!
                            \begin{array}{c}
                                      #1\\
                                      #2
                            \end{array}\!
                       \right\}}

\begin{document}

\title{Isoscalar Giant Dipole Resonances in Spherical Nuclei -  Macroscopic Description}

\author{\c Serban Mi\c sicu}
\email{misicu@theor1.theory.nipne.ro}
\homepage{http://www.theory.nipne.ro/~Emisicu/}
\affiliation{Department for Theoretical Physics, 
National Institute for Nuclear Physics, Bucharest,
P.O.Box MG6, Romania}


\begin{abstract}The properties of the Isoscalar Giant Dipole Resonance (ISGDR) and 
its electromagnetic structure are investigated within a semiclassical nuclear Fermi-Fluid dynamical 
approach. 
Microscopical calculations pointing to a ISGDR distribution splitted into two main broad structures is confirmed within the presented macroscopic approach by the occurence of a 
''low-lying'' and a ''high-lying'' state, that are nothing else than the 
first two overtones of the same resonance. Macroscopically they are pictured 
as a combination of compressional and vortical nuclear flows. 
In the second part of the paper the electromagnetic structure of the
ISGDR, relevant for reactions with inelastically scattered electrons, and the 
relation between the vorticity and the toroidal dipole moment is analyzed.
The relative strengths of the compresional and vortical collective currents 
is evaluated by means of electron-scattering sum-rules. 
\end{abstract}

\maketitle

\section{Introduction}

Recently the College Station group \cite{clark01} reported experimental 
results on the isoscalar $E$1 strengths in three proton magical nuclei  
($^{90}$Zr, $^{116}$Sn and $^{208}$Pb) using inelastic scattering of 
240 MeV $\alpha$ particles at small angles.
The authors concluded that the isoscalar $E1$ strength distribution in each 
nucleus  is shared mainly between two components, one located at low energy and
and another one at higher energy.
In a subsequent publication this group presented new data 
on the ISGDR \cite{young04}. 
For $^{116}$Sn, $^{144}$Sm and $^{208}$Pb the low-energy peak fall in the interval $(1.71-1.92)\hbar\omega$ whereas the high-energy peak lays between $3\hbar\omega$ and $3.2\hbar\omega$. The upper component covers approximately 3 times more of the 
energy-weighted sum rule compared to the lower component. 
Similar values for the two peaks for $^{208}$Pb are given in \cite{uch03} :
$1.80\hbar\omega$ and  $3.25\hbar\omega$.
Previously Morsch {\it et al.} \cite{morsch83} found for the
high-lying component in $^{208}$Pb a centroid of $21.3\pm 0.8 $MeV
which corresponds to $(3.15\pm 0.12)\hbar\omega$. 
Therefore, experimentally, the lower-energy component has a value very  
to the IVGDR centroid which lays around $\approx 2\hbar\omega$, whereas
the higher-energy component is located in the same region as the electric 
octupole resonance, i.e. $\approx 3\hbar\omega$. 

On the theoretical side there have been numerous studies aiming to disclose 
the features of these exotic modes. The usually accepted macoscopical picture of the ISGDR is a "hydrodynamical density oscillation" in which the volume of the nucleus
remains constant and the proton-neutron fluid oscillates in phase back and forth through the nucleus in the form of a compression mode \cite{deal73,har81}. 
Microscopical calculations using strengths associated with the non-isotropic 
compression, namely the "dipole squeezing" operator 
$D=\sum_i r_i^3 Y_{1\mu}({\hat r}_i)$  are stressing the importance of the high-lying ISGDR 
($\approx 3\hbar\omega$)  \cite{wamb78,vansag81,hamsag98}. They are also predicting a rather fragmented peak at smaller energies  ($\approx 2\hbar\omega$).  
\cite{dum83,col00,shlo02,gor04}.  A similar, bimodal structure, was obtained within the fully
consistent relativistic Hartree-Fock plus RPA framework \cite{vret00} 

The first macroscopically based models of electric resonances, were primarly concerned to describe the ISGDR as density fluctuations of a nuclear liquid drop. In an
apparently overlooked paper, available only in german \cite{woeste52}, and published 
a couple of years after the emergency of the incompressible fluid model of the IVGDR, for the first time the isoscalar dipole eigenfrequencies of a spherical nucleus were determined. 
In our days, the investigations of the College Station - Kyiw group \cite{kol01}, 
pointed out that considering only compressional components in the velocity field 
and neglecting the relaxation effects within the nuclear-fluid dynamics, leads to an overestimation of the energies of the $1^-$ resonances with respect to the experimental values.

Other macroscopical approaches aimed at the description of the giant resonances 
(including the 1$^-$, $T=0$ state) by allowing for vortical components along with or
without the compressional(incompressional) of the velocity field. 
Deriving conservation equations, such as continuity equation, and equations of motion
such as the Navier-Stokes or Lam\'e it was shown that the macroscopic velocity
field admits also shear (transverse) components \cite{holz78}.  
In ref.\cite{sem81}, after some simplifications compared to \cite{holz78}, that we are going to dismiss in the present study, an isoscalar $1^-$ state of pure vortical character was derived. Substracting the center-of-mass motion the associated velocity field  reads
\beq
{\bd{v}_\mu}^{\rm tor}(r,\theta,\phi)
=\nabla\times\nabla\times\bd{r}\left (r^3-\frac{5}{3}r\langle r^2\rangle \right)
Y_{1\mu}(\theta,\phi)
\eeq
Most important, in this study, for the first time a connection to the toroidal class 
of electromagnetic multipole moments was done introduced earlier by Dubovik and Cheshkov \cite{dubche74}. 
The theoretical search for a vortex-like isoscalar dipole electric excitation
associated to the toroidal dipole moment(''dipole torus mode'') was continued in the nuclear-fluid dynamics frame \cite{balmik88,basmis93}. Ref.\cite{basmis93} substantiated the elastic character of this mode, since a nucleus without shearing properties cannot
withstand transverse-like oscillations, and evaluate for the first time the
$(e,e^\prime)$ form factors corresponding to the excitation of this isoscalar
$1^-$ resonance. The radial part of the transverse electric form-factor corresponding to the electro-excitation of the dipole torus mode was found to vary like $j_3(qr)/qr$.

Apart from the quest of toroidal nature of the ISGDR  
there are also other issues related to the enhancement of these electromagnetic transitions
for other types of collective electric excitations. The electric dipole spin waves
were identified to have non vanishing magnetization-dependent part of the toroidal dipole
operator in ref.\cite{mis95a}. 
In \cite{mis95b} and \cite{mikh96} the fingerprints of the dipole toroidal
moments in the electromagnetic properties of nuclear rotational states were examined
for the first time in the literature.
In \cite{mikh96} it was inferred that the strong deviations from the 
predictions of the adiabatic theory for the absolute values of $E$1-transitions
in the Coulomb excitation of $^{226}$Ra are related to the enhancement of toroidal 
transitions  between the ground state and the lowest negative parity band.

The revival of the interest on the role played by toroidal moments in the excitation
of the isoscalar dipole resonance was caused by a recent publication 
\cite{cro-ring01} that aimed to evaluate the $E$1 strength distribution in 
spherical nuclei where the RMF formalism + RPA calculations were previously
unable to provide a satisfactory agreeement with the experimental data on 
the positions of the ISGDR resonances. 
Using the toroidal dipole operator corrected for the c.m. motion 
instead of the squeezing operator, broad resonant peaks were assigned in the low-
($\leq 2\hbar\omega$) and high-energy ($> 3\hbar\omega$) regions of the
strength distribution for $^{208}$Pb.

Commenting on the results reported in \cite{cro-ring01}, the short-note 
\cite{misbas02} stressed the fact that the low-lying vortical mode, as inferred
from macroscopic calculations \cite{sem81,balmik88,basmis93} is different from the 
so-called ''pigmy resonance'' which lays in the vicinity of 1$\hbar\omega$, and that the centroids provided by the nuclear-fluid approaches \cite{balmik88,basmis93} are still 
providing a qualitative good agreement with the $(\alpha,\alpha^\prime)$ scattering data. 
In this respect the merit of \cite{cro-ring01} is that it offers for the first time in the 
literature a microscopical calculation of the toroidal content of ISGDR states and confirms the connection already established by macroscopic models between this electromagnetic characteristic and 
vorticity. Moreover, and this will be an important point in our present work, although not explicitly stated in the body of ref.\cite{cro-ring01} but rather inferred from Fig.2, it indicated that also for the high-lying states there is a more or less important value of the toroidal strengths which implies that these excitations are not purely compressional but may contain also significative vortical admixtures.

The nature of collective flows in a range of excitation energy below 20 MeV for
$^{208}$Pb was more closely approached in \cite{ryez02}, where calculations within the 
quasiparicle  phonon model are pointing to strong vorticity below 2$\hbar\omega$
for the entire electric dipole response, not only the isoscalar one. 

In a subsequent publication \cite{kvasil03} the nuclear electric isoscalar
dipole response was studied within the RPA formalism including in the 
dipole part of the separable interaction simultaneously dipole-dipole 
($F_{\lam\mu}=rY_{1\mu}$) and compressional dipole(squeezing) 
($F_{\lam\mu}=r^3Y_{1\mu}$) fields. The authors infered that the isoscalar 
dipole response is less sensitive to these type of interactions and more to 
the single particle structure. The strength function for  what they called 
the squeezed dipole mode, 
displays two main  broad and fragmented peaks : a low-lying, which depends on the 
coupling constants of the Nilsson potential and has a centroid ranging between 8 and 11 MeV and a high-lying one with centroids ranging in the interval 21-23 MeV.

The present paper aims to a description of the properties of the ISGDR within a 
macroscopic model based on the nuclear Fermi-Fluid picture, where along with density fluctations, transverse components of
the collective velocity field are included. 
The main purpose is to determine the share of compression and vorticity
flows in the states building the isoscalar dipole electric response.
The electromagnetic structure of the ISGDR is studied for arbitrary momentum transfer
by using the multipole parametrization of charges and currents of ref.\cite{dubche74}. 
The relation between the transition electric dipole moments and the transition 
vorticity is discussed. 
Finally we introduce inelastic electron scatterig sum-rules in order to assess
the strengths of the states building the ISGDR with varying momentum transfer. 
  
\section{Nuclear-Fluid Dynamic Approach}

A macroscopic approach which goes partially on the lines already developed 
in a previous publication \cite{basmis93} is adopted. However a few strong 
amendments are performed :
\begin{itemize}
\item full $k$-content in the radial part of the collective field 
\item a compressional elastic constant different from the shear elastic constant ($\lambda_{\rm Lame} \neq\mu_{\rm Lame}$) that draw it nearer to other Fermi-Fluid dynamical approaches \cite{kol01,holz78,wong6}.
\end{itemize}

The procedure consists in taking moments of the  Boltzmann equation, i.e.
to integrate it in the momentum space with weights 1, $p_i$, $p_ip_j$, etc..
The first two moments  provide the continuity
and the Navier-Stokes equation in view of their similar form to the well-known
equations known from Hydrodynamics. 
\beq
\frac{\partial \rho(\bd{r},t)}{\partial t}+\nabla\cdot(\rho(\bd{r},t){\bd{\dot{u}}})
=0
\label{continuity}
\eeq
\beq
\frac{\partial}{\partial t}({\rho(\bd{r},t) u_i})+\sum_{j=1}^3\frac{\partial P_{ij}}{\partial x_j}=0
\label{navier}
\eeq
where $\rho$ is the mass density, which is supposed to be of the sharp-edge type
in the present work, $\bd{u}$ is the collective field (which vanish in the ground state), whereas $P_{ij}$ are stress tensor components.  
To these equations the linearization procedure is applied 
\beq
\rho(\bd{r},t)=\rho_0+\delta\rho(\bd{r},t),~~~\bd{u}=\delta\dot{\bd{s}},~~~
P_{ij}=p\delta_{ij}+\delta\sigma_{ij}
\label{linearization1}
\eeq
In the second equation of (\ref{linearization1}) the displacement field, 
$\delta{\bd{s}}$, was introduced. The stress
tensor is splitted in a diagonal part (normal pressure)
\beq
p=-\frac{1}{9m}K\rho_0 \dot{\cal D}
\label{pressure}
\eeq 
and a non-diagonal one (associated to the shear)
\beq
\delta\sigma_{ij}=-{\frac{4}{5m}}\rho_{0}\epsilon_F
\left( \varepsilon_{ij}-\frac{1}{3}\delta_{ij}{\cal D} \right) 
\label{shear}
\eeq
In the above two formulas $K$ is the incompressibility coefficient
of nuclear matter and $\epsilon_F$ is the Fermi energy. 
The scalar function 
 $${\cal D}\equiv\nabla\cdot\delta\bd{s}$$     
describes the compressibility  of the displacement 
field $\delta\bd{s}$ and $\varepsilon_{ij}$ are the components
of the dyadic strain tensor \cite{som78},
$$\widehat{\bd{\varepsilon}}=\oh\left( \nabla\bd{s}+\bd{s}\nabla\right) $$

After linearizing the equations of motion we get 
\beqa
\delta\dot\rho&=&\rho_0\dot{\cal D}  \\
\rho_0\delta\ddot{\bd{s}}&=&(\lam_{\rm Lame}+2\mu_{\rm Lame})
\nabla(\nabla{\cal D})-
2\mu_{\rm Lame}\nabla\times\bd{\omega}
\label{lame_eq}
\eeqa
where, like in Hydrodynamics (\cite{som78}, p.115), by $\bd{\omega}$ we denote the vorticity vector which is proportional to the curl of the displacement field
$$\bd{\omega}\equiv{\oh}\nabla\times\delta\bd{s}$$

The equation of motion (\ref{lame_eq}) is identical to the
Lam{\'e} equation (\cite{som78}, p.60 and 94) known in the Mechanics
of deformable continua, where the Lam{\'e} elastic coefficients are
provided by the properties of the nuclear Fermi gas \cite{wong6}
$$ \lam_{\rm Lame}=\frac{n_0 K}{9}-\frac{4}{15}n_0\varepsilon_F, ~~~~~~
\mu_{\rm Lame}=\frac{2}{5}n_0\varepsilon_F$$ 
 For the incompressibility coefficient $K$
we use the fact that the excitation energy of the Isoscalar Giant
Monopole Resonance (ISGMR), which is an isotropic volume oscillation, can be
related to the compressibility of nuclei  
\cite{bla80}, i.e.  
$$E_{\rm ISGMR}=\left ( \frac{\hbar^2 K}{m\langle r^2 \rangle_0}\right )^{1/2}
\approx 82\cdot A^{-1/3}$$

In what follows a fundamental
theorem of vector analysis is used (\cite{som78}, p.131) which states that 
every  continous vector field $\bd{V}$, which, together with its derivative 
falls to 0 at large distances can be decomposed into a  divergenceless part 
$\bd{V}_{\perp}(\nabla\cdot\bd{V}_{\perp}=0)$ and a curless part
$\bd{V}_{\parallel}(\nabla\times\bd{V}_{\parallel}=0)$. 

Then eq.(\ref{lame_eq}) separates in a equation for the compresibility 
(${\cal D}$) and another one for the vorticity ($\bd{\cal \omega}$) 
which  describes the degree of shear of the displacement field
for the case of an elastic body.   
\beq
\ddot{\cal D}=c_L^2\Delta{\cal D},~~~~
\ddot{\bd{\omega}}=c_T^2\Delta{\bd{\omega}}
\eeq
where $c_L=\sqrt{\lam_{\rm Lame}+2\mu_{\rm Lame}/\rho_0}$ and 
$c_T=\sqrt{\mu_{\rm Lame}/\rho_0}$ are the
propagation velocities of the longitudinal(compressional) and transversal
(shear) elastic waves in nuclear matter \cite{wong6}.

Assuming an harmonic variation in time of the fluctuating parts of the 
density and the displacement field, i.e.
$$\delta\rho(\bd{r},t)=\rho(\bd{r})e^{i\Omega t},~~~
  \delta\bd{s}(\bd{r},t)=\delta\bd{s}(\bd{r})e^{i\Omega t}$$ 
the compresibility and the vorticity  are found to satisfy the scalar and vector 
Helmholz equation respectively
(HE)
\beq
\left ( \Delta+\macol{k_L^2}{k_T^2} \right )
\macol{\cal D}{\bd{\omega}}=0
\label{helmhol}
\eeq
corresponding to the wave-numbers $k_{L,T}=\Omega/c_{L,T}$.
In seismology the compresional wave is called the $P$ wave
and the transverse wave the $S$ wave \cite{ben81}. The $S$ in its turn
has two components : the $SH$ wave (known as the ''poloidal'' in Hydrodynamics
or ''transverse electric'' in Electrodynamics) and the $SV$ wave (''torsional'' 
or ''magnetic'').
For a nucleus with a sharp edge one adopts a spherical geometry and the radial part
is given by spherical Bessel functions whereas the angular part can be written 
in terms of spherical harmonic vectors. Details can be found in the 
appendix or in the literature \cite{morse53}.
Next we disregard the torsional component which is related to
magnetic excitations \cite{bastmag} and consider only axial-symmetric
displacement fields ($\mu=0$). 
We then have for the longitudinal and poloidal components of the displacement
field the expressions (\ref{app-long}) and (\ref{app-tran}) derived in the
appendix.
These expressions have to be further corrected in order 
account for the center-of-mass motion. Like in a preceeding paper 
\cite{basmis93} the translational invariance of the collective velocity 
field results from the condition that the center-of-mass $\bd{R}_{C.M.}$
is at rest
\beq
\delta\bd{R}_{C.M.}=\frac{\int d\bd{r}\rho(\bd{r},t)\delta\bd{s}}{\int d\bd{r}\rho(\bd{r},t)}=0
\label{cmcorr}
\eeq
Thus in the dipole case ($\lambda=1$) the  longitudinal and transverse  
displacement fields are
\beq
\delta\bd{s}_L(\bd{r},t)=\frac{1}{\sqrt{3}}a\left [ \left( j_0(k_L r)
-\frac{3}{k_L R_0 }j_1(k_L R_0)\right)\bd{Y}_{10}^{0}(\theta,\phi)
+\sqrt{2}j_2(k_L r)\bd{Y}_{12}^{0}(\theta,\phi)\right ]
\label{slcorr}
\eeq
\beq
\delta\bd{s}_T(\bd{r},t)=-\frac{1}{\sqrt{3}}b\left [\sqrt{2} \left( j_0(k_T r)
-\frac{3}{k_T R_0}j_1(k_T R_0)\right)\bd{Y}_{10}^{0}(\theta,\phi)
-j_2(k_T r)\bd{Y}_{12}^{0}(\theta,\phi)\right ]
\label{stcorr}
\eeq
The corresponding corrected expression for the density fluctuation results by applying
the continuity equation in (\ref{cmcorr}) followed by the substitution of the 
longitudinal diplacement field (\ref{slcorr}). The integral relation between the corrected
expression of the density fluctuation and the longitudinal displacement field reads
\beq
  \int d\bd{r}~\bd{r}\delta\rho=\rho_0\int d\bd{r}~\nabla\times(\delta\bd{s}_L\times\bd{r})
\eeq  
Eventually, we get for the density fluctuation an expression identical to the one 
derived in \cite{kol01}
\beq
\delta\rho=a\rho_0\left ( j_1(k_L r)\Theta(R_0-r)
-\frac{1}{k_L}j_2(k_L R_0)\delta(R_0-r)\right ){Y}_{10}(\theta,\phi)
\label{denscorr}
\eeq

In order to derive the longitudinal and transverse wave-numbers $k_L$ and $k_T$
and the constants $a$ and $b$, multiplying the displacements fields, 
boundary conditions for the force acting on the free surface of the nucleus 
have to be imposed. This force is obtained by 
projecting the dyadic stress tensor on the normal unit vector to the surface. 
\beq
     \bd{F}=\widehat{\bd{P}}\cdot\bd{e}_r
\eeq
Usually two types of boundary conditions are employed depending on what 
assumption has been made for the surface.  Two kinds of bounding nuclear surfaces
are distinguished for sharp-edge distributions : {\em rigid surfaces} on which no slip
occurs (used in the liquid drop-model to determine the ''surfon'' eigenvalues
or in the hydrodynamic model of giant resonances \cite{gre70} to determine the 
''gion'' eigenvalues) and {\em free surfaces} on which no tangential stresses act. 
If we would assume a rigid surface then we would end up with a density fluctuation,
and thus also with a velocity field, containing admixtures of the center-of-mass(c.m.)
motion. Actually the expression for the fluctuation density (\ref{denscorr}), corrected for the c.m. motion, is compatible with the assumption of a free surface.
The boundary conditions on a free surface impose that the force fulfill the following two conditions \cite{wong6,holz81}:
\beq
\left.\bd{e}_r\cdot \bd{F}\right |_{r=R_0}=\left.P_{rr}\right
|_{r=R_0}=0
~~~~
\left.\bd{e}_r\times \bd{F}\right |_{r=R_0}=
\left.(\bd{e}_{\phi}P_{r\theta}-\bd{e}_{\theta}P_{r\phi})
\right|_{r=R_0}=0
\label{bound-cond}
\eeq
The above equations provides an infinity of eigenvibrations but 
for the study of giant resonances only the first few are relevant. 
The boundary condition  (\ref{bound-cond}) allows also the
determination of the ratio $b_n/a_n$ which gives
the admixture between the compressional(longitudinal) and vortical(transverse)
field in a given state $n$
$$r_n\equiv\frac{b_n}{a_n}=-\left.\frac{P_{rr}^{\rm Longitudinal}}
{P_{rr}^{\rm Transversal}}\right|_{r=R_0}$$ 

In table \ref{tabela1} we list the first 4 roots(overtones)  of the boundary condition 
plus the ratio of the transversal-to-longitudinal weights.

\begin{table}
\begin{center}
\begin{tabular}{cccccc}
\hline
\hline
Overtone($n$)&$k_L^{(n)}$(fm$^{-1}$)& $k_L^{(n)}R_0$ &$\hbar\Omega_n$ (MeV)& 
$\hbar\Omega_n/\hbar\omega_0$ & $r_n\equiv b_n/a_n$\\
\hline
1 &2.05 &  3.05 & 11.56  &  1.67 &  1.94\\
2 &3.93 &  5.86 & 22.19  &  3.21 & -1.19\\
3 &5.05 &  7.53 & 28.53  &  4.12 &  6.09\\
4 &7.09 & 10.57 & 40.03  &  5.78 & -4.03\\
\hline
\hline
\end{tabular}
\caption{The first 4 overtones of provided by the eigenvalues of the
boundary condition (\ref{bound-cond}) and the 
vorticity/compressibility ratio for $^{208}$Pb}
\label{tabela1}
\end{center}
\end{table}

\begin{figure}
\centering
\mbox{\subfigure{\epsfig{figure=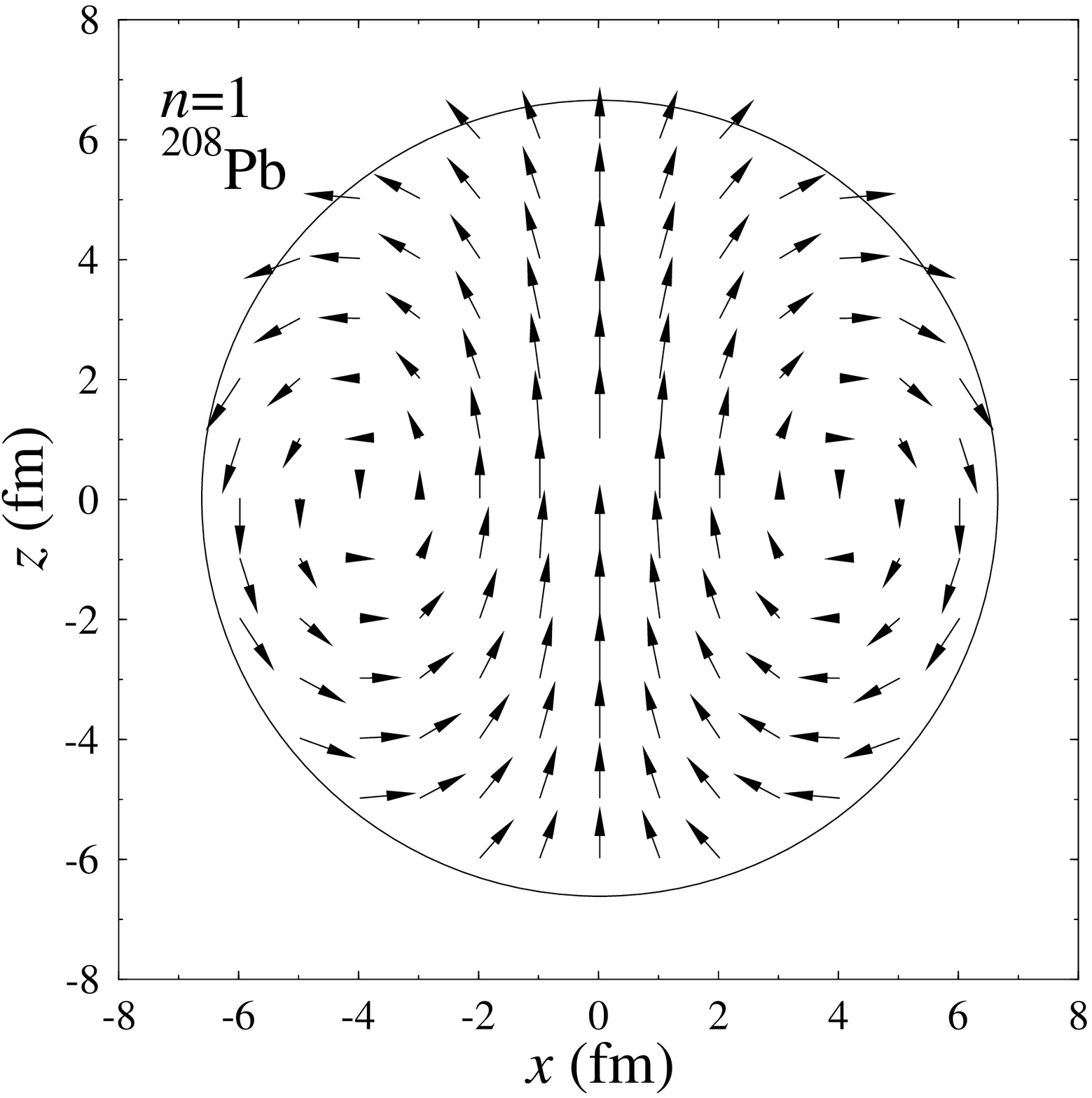,width=0.5\textwidth}}
      \subfigure{\epsfig{figure=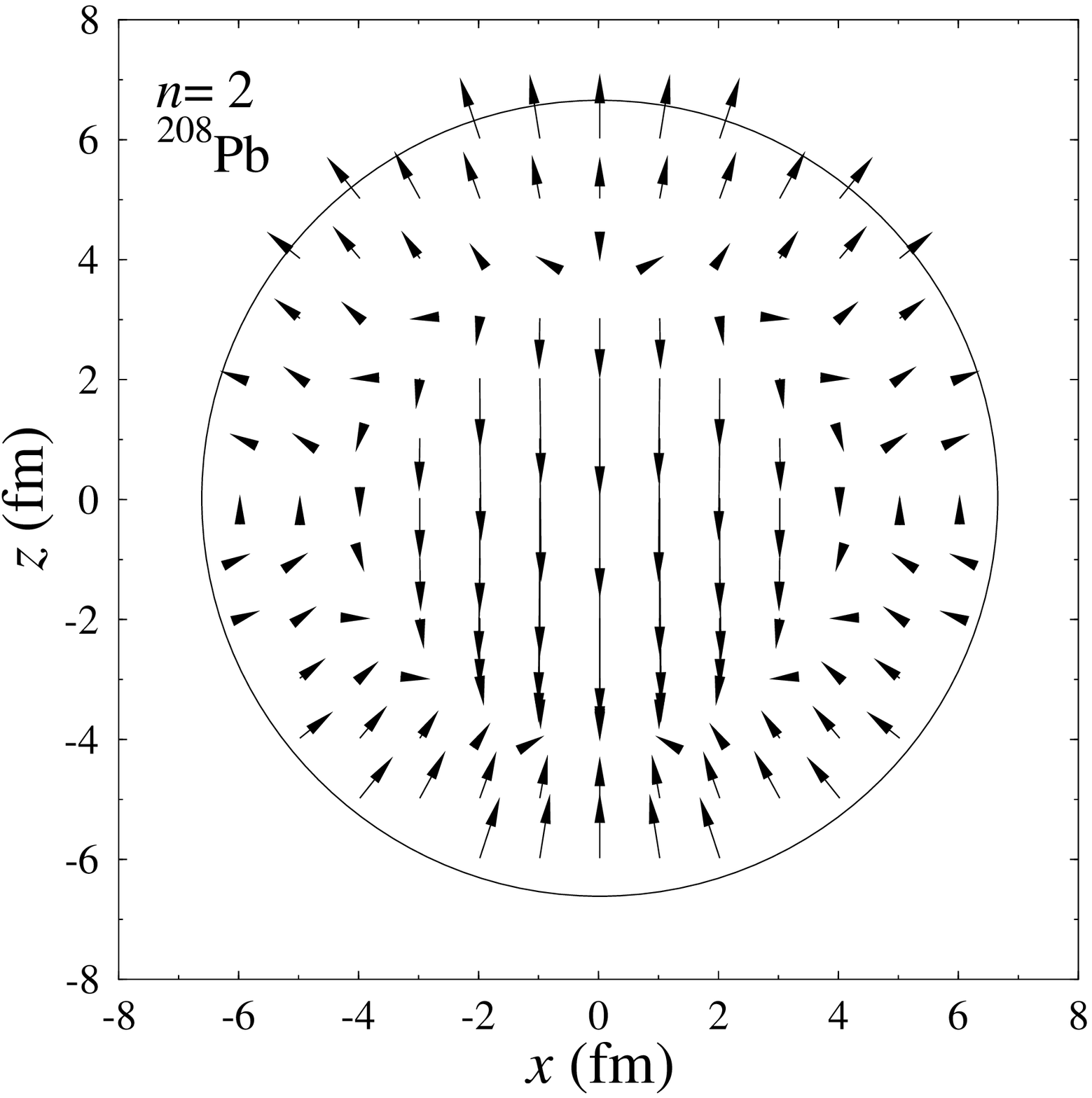,width=0.5\textwidth}}
     }
\mbox{\subfigure{\epsfig{figure=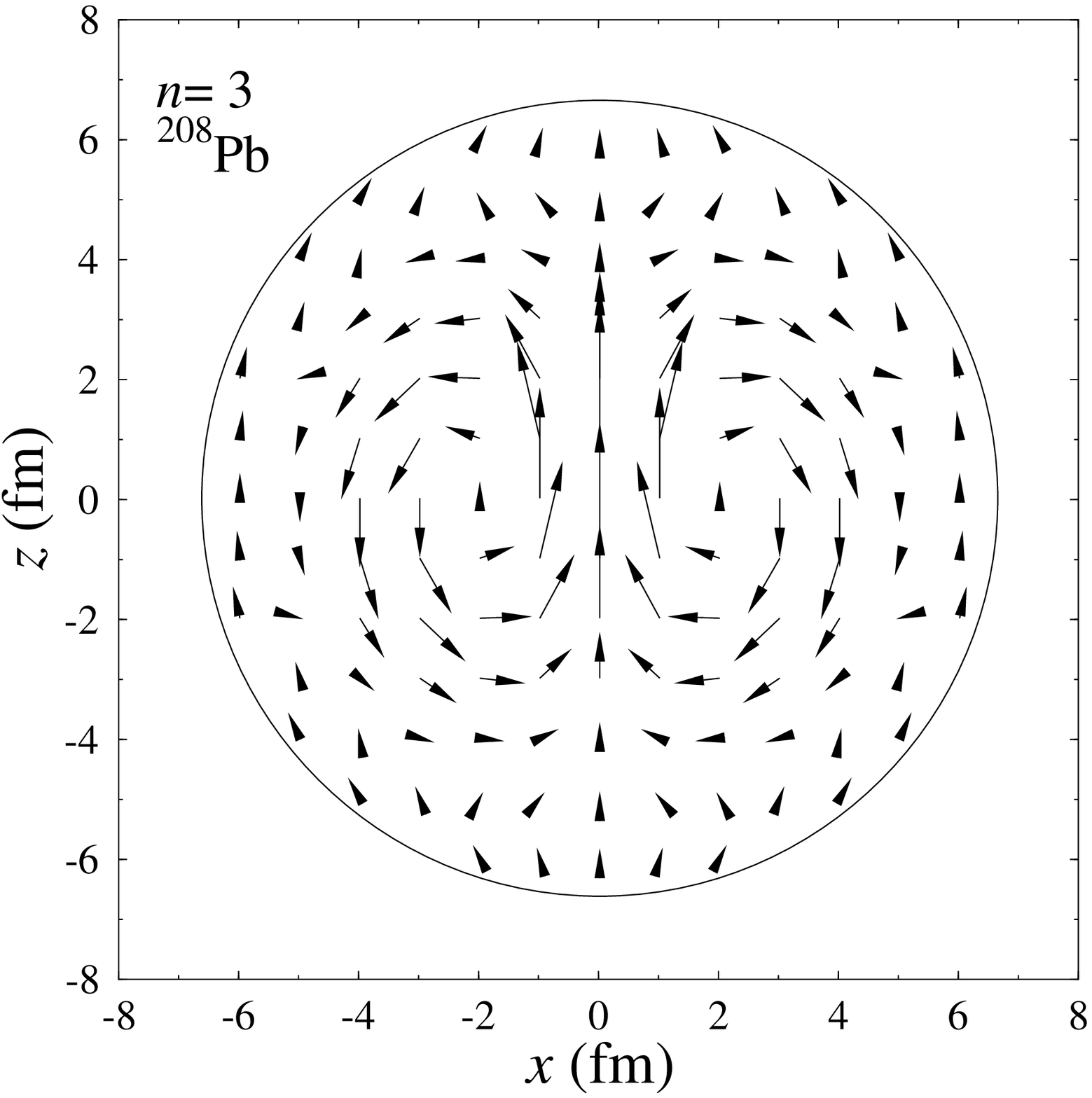,width=0.5\textwidth}}
      \subfigure{\epsfig{figure=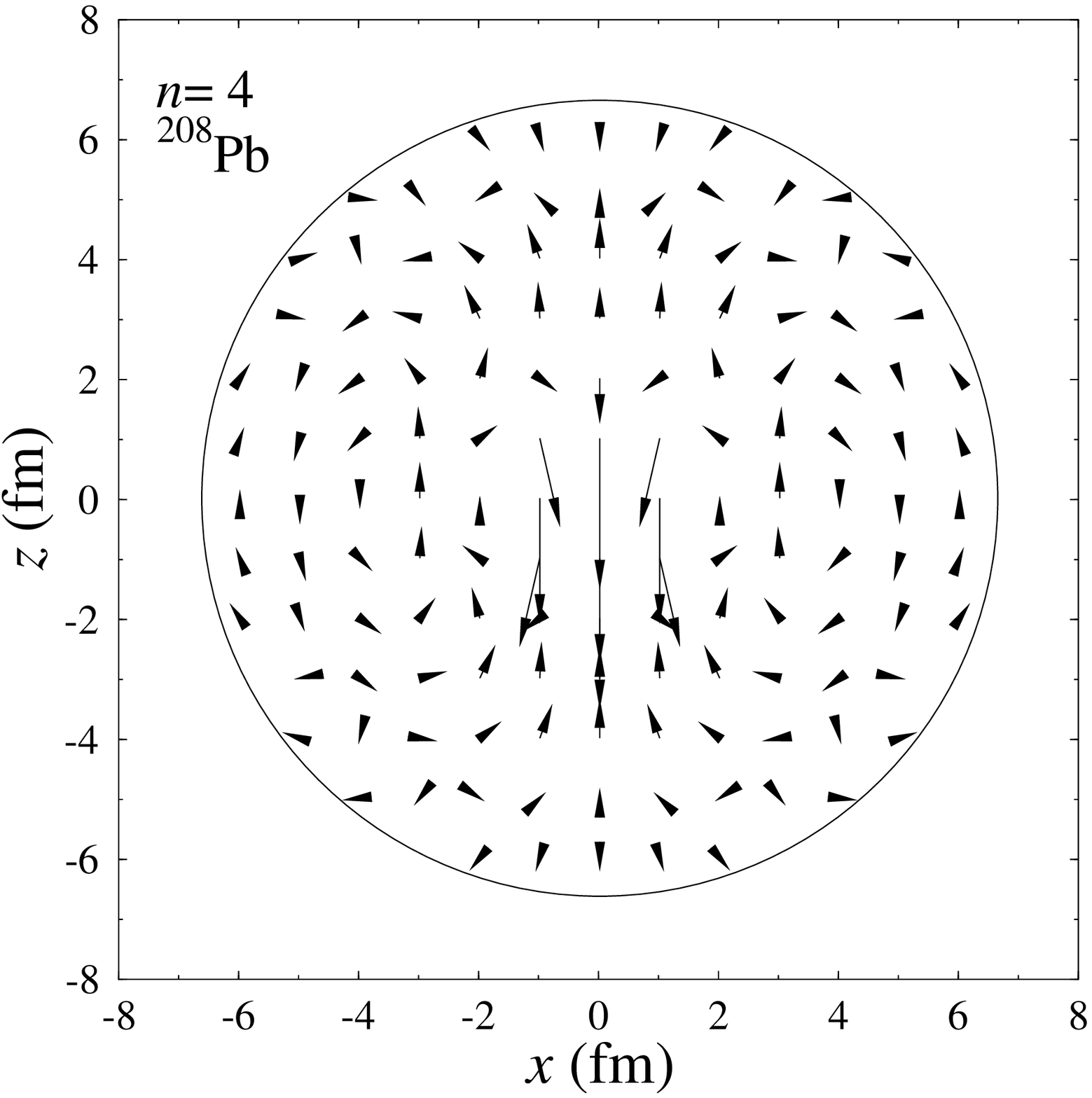,width=0.5\textwidth}}
      }
\caption{Flow lines corresponding to the first four overtones of the ISGDR in 
$^{208}$Pb.}
\label{flows}
\end{figure}

The first three eigenfrequencies for the dipole density fluctuations as derived
in the pioneering work of Woeste \cite{woeste52} have values very close to the 
density-vorticity waves listed in the above table :
1.78$~\hbar\omega$, 3.02$~\hbar\omega$ and 4.22$~\hbar\omega$ compared to
1.67$~\hbar\omega$, 3.21$~\hbar\omega$ and 4.12$~\hbar\omega$.
Also the estimation for the $T=0, L=1^-$ vortex mode from \cite{sem81} 
(1.7$\hbar\omega$) is very close to the value derived in the present paper. 
In a previous paper dedicated to ISGDR \cite{basmis93} together with collaborators we constrained the collective velocity field to admit purely vortical flows and we considered the low-lying response of the
Fermi liquid, whereas in the present approach the full momentum content is taken into account.
Therefore, although compressional components are occuring, the first overtone has a rather vortical character as can be easily inferred from the 
upper left panel of Fig.\ref{flows}, and displays the typical {\em Hill-vortex} pattern as already mentioned in \cite{sem81,balmik88,basmis93} : Nuclear matter flows around a 
vortex ring situated in the equatorial plane of the  nucleus.
Other works, based like \cite{woeste52} on the compressible and irrotational nuclear liquid
drop, e.g. \cite{boh75}, fail to observe the first overtone. In such approaches the necessity to correct the density fluctuation for the c.m. motion was not anticipated.
For the second overtone the compressional and vortical flows have almost an equal importance, which contradicts the entrenched picture
of a compressional mode around 3$\hbar\omega$. However there is to the date no direct experimental indication on such a macroscopic property of this higher lying isoscalar 
dipole resonance, and therefore the result that we report on this mode should not be excluded
from debate. To a certain extent the flow pattern of this mode (see upper right panel of 
Fig.\ref{flows}) presents typical characteristics of a compressional mode : the concentration of nuclear matter flow 
inside the southern emisphere and the depletion inside the northern emisphere; in the same time an opposite behavior of the density fluctuation is manifested at the north and south poles.
For the third and fourth overtones (see lower left and right panels of Fig.\ref{flows}) the
flow is predominantly vortical.
Although difficult to disclose from the flow patterns, the vortex structure becomes
more intricate in the sense that instead of one vortex ring as was the case for the
first overtone, we deal with two rings ($n=3$ overtone) and respectively three rings ($n=4$
overtone) of lower intensity. Note that two succesive rings have opposite rotational flows.

A quantitative way to asses the role of compressional and vortical flows is to introduce,
following \cite{holz81} the orientation averaged values of the collective velocity field divergence 
$$\langle {\cal D}\rangle\equiv\left( \int{\cal D}^2d\Omega\right )^\oh$$ 
and vorticity 
$$\langle \bd{\omega}\rangle \equiv \left (\int\bd{\omega}^2d\Omega\right )^\oh$$

\begin{figure}[b]
\centering
\mbox{\subfigure{\epsfig{figure=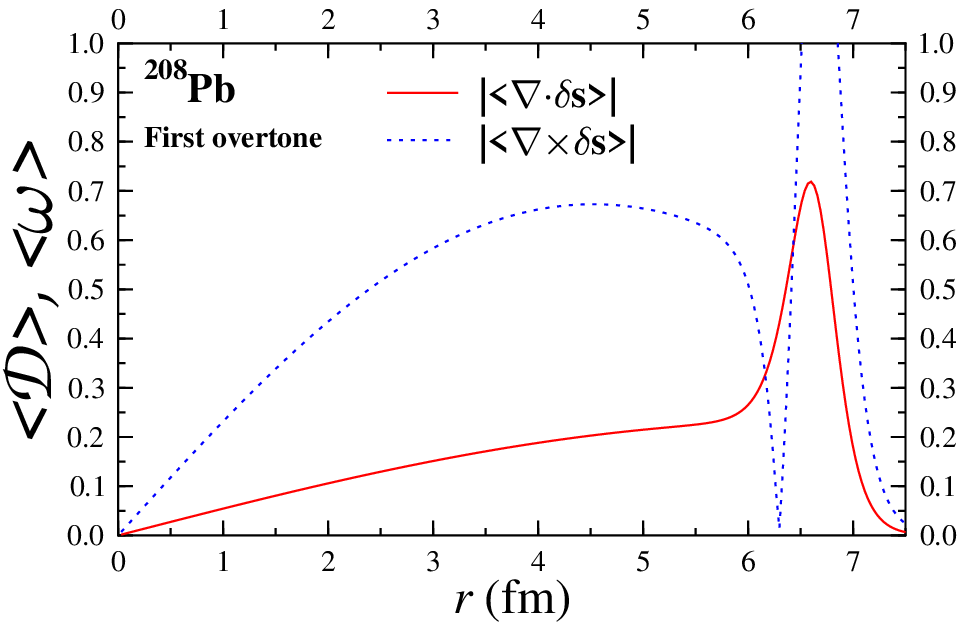,width=0.5\textwidth}}
      \subfigure{\epsfig{figure=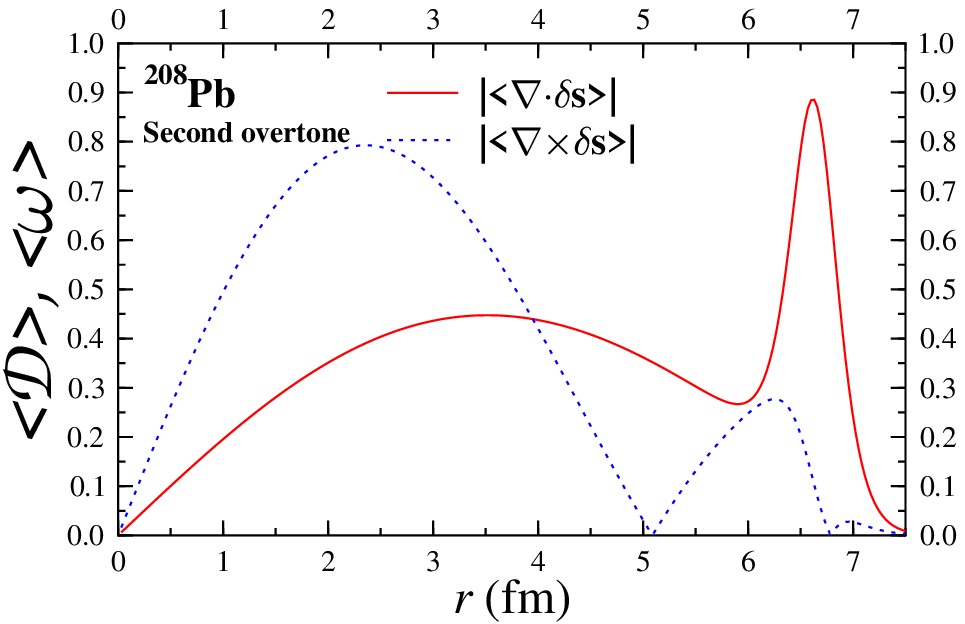,width=0.5\textwidth}}
     }
\mbox{\subfigure{\epsfig{figure=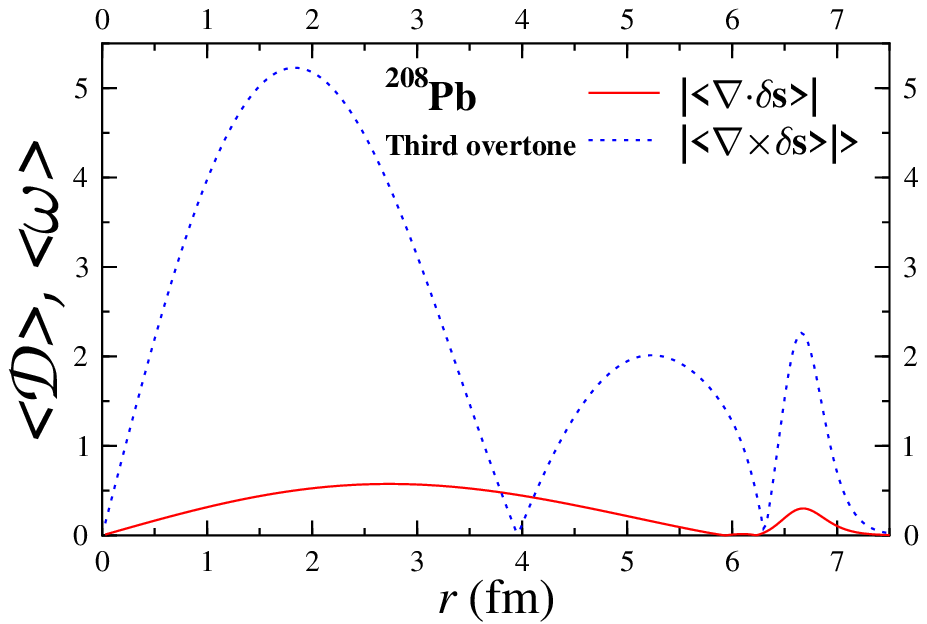,width=0.5\textwidth}}
      \subfigure{\epsfig{figure=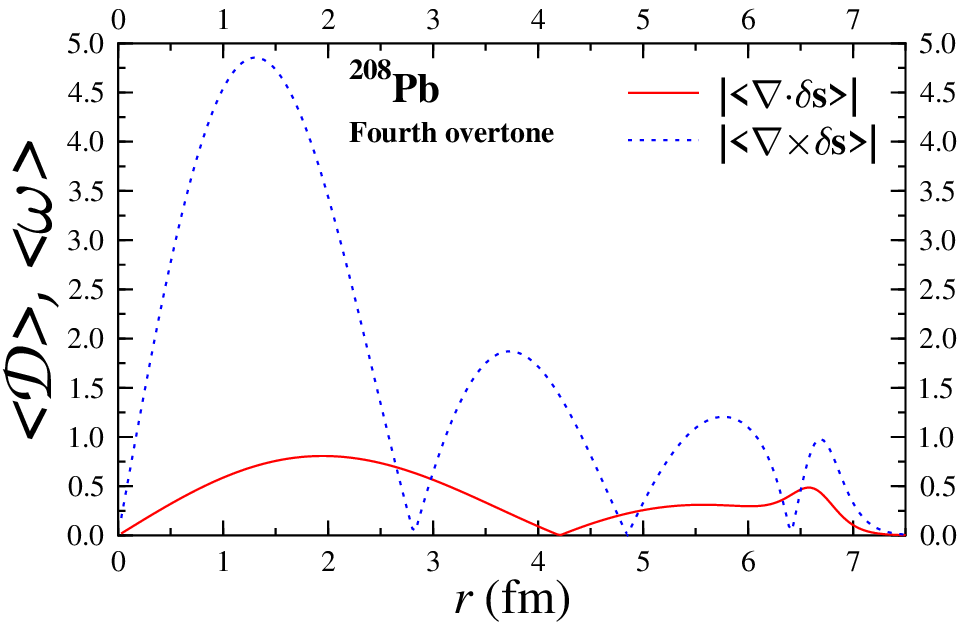,width=0.5\textwidth}}
      }
\caption{Compresibility and vorticity of the first four overtones of the 
ISGDR in $^{208}$Pb.}
\label{compvor}
\end{figure}

These two quantities are displayed in Fig.\ref{compvor}.
In order to avoid the awkward effect on these two radial functions near the surface, 
which is due to the sharp-edge distribution,  (see ref.\cite{holz81}), 
the curves drawn in Fig.\ref{compvor} were computed by assuming a diffuse density distribution. 
Consequently an additional peak in both $\langle {\cal D}\rangle$ and 
$\langle \bd{\omega}\rangle$ occurs in the surface region. In what concerns the vorticity we remark for the first overtone that 
the vorticity attains a maximum at approximately $R_0/\sqrt{2}$, which corresponds to the
critical points of the Hill vortex, a fact already pointed out in \cite{basmis93}.
Inside the nucleus the compressibility increases almost linearly with the radius and is less
important than the vorticity, thus confirming the previous assignment of this collective state as dipole torus mode \cite{sem81,balmik88,basmis93}. When the overtone number increases the 
vortex with the largest strength migrates towards the center of the nucleus, and new ring-like vortices occur at larger radii. For the overtones with $n\geq 2$ the compressibility developes 
maxima inside the nuclear sphere and it plays a dominant role only for the $n=2$ overtone
(the claimed 3$\hbar\omega$ "compression mode") in the vicinity of the nuclear surface. 
The mixed (compressional+vortical) character of the $n=2$ state can be also 
inferred from the self-consistent RPA calculations with Skyrme type interaction
performed in an old study \cite{serr83} as well in the relativistic mean-field approach from \cite{vret00}.

The procedure to quantize a continuum system, described by the 
equation of continuity and the equation of motion (\ref{lame_eq})
is to expand $\delta\rho$ and $\delta\bd{s}$ in normal coordinates
\cite{feenberg69} :
\beq
\delta\rho(\bd{r},t)=\sum_n{\cal \rho}_n(\bd{r})\alpha_n(t),~~~~~  
\delta\bd{s}(\bd{r},t)=\sum_n\bd{s}_n(\bd{r})\alpha_n(t)  
\eeq
The above sums are running after the values of $k_L(k_T)$ allowed
by the boundary condition (\ref{bound-cond}), i.e. after the overtones $n$.
For the expression of the kinetic energy in the newly introduced
collective coordinates $\alpha_n$ we have
\beq
T=\oh\rho_0\int~d\bd{r}|\delta\dot{\bd{s}}(\bd{r},t)|^2\equiv
\oh\sum_n B_n|\dot{\alpha}_n|^2
\eeq 
For the mass inertia parameter we derive the expression
\beqa
B_n=\frac{3mA}{4\pi}\left\{ \left [ \oh\left (j_0^2(k_L^{(n)}R_0)+j_1^2(k_L^{(n)}R_0)\right )
-\frac{1}{2k_L^{(n)}R_0}j_1(k_L^{(n)}R_0)\left(j_0(k_L^{(n)}R_0)
+\frac{6}{k_L^{(n)}R_0}j_1(k_L^{(n)}R_0)\right )
\right ] \right.\nn\\
\left. 
+r_n^2\left [ \oh\left (j_0^2(k_T^{(n)}R_0)+j_1^2(k_T^{(n)}R_0)\right )
-\frac{1}{2k_T^{(n)}R_0}j_1(k_T^{(n)}R_0)\left ( j_0(k_T^{(n)}R_0)+\frac{6}{k_T^{(n)}R_0}j_1(k_T^{(n)}R_0)\right )
\right ]
\right\}\nn\\
\eeqa
The stiffness coefficient associated to an oscillation of degree $n$
is simply
$$C_n=\Omega_n^2B_n$$ 
and the quantized form of the energy can be obtained by introducing 
the creation and annihilation dipole ''gions'' \cite{gre70}
$$
\hat{d}_n^+=\left ( \frac{\Omega_nB_n}{2\hbar}\right )^\oh
\left ( \alpha_n-\frac{i}{\Omega_n}\dot{\alpha}_n\right ),~~~~~~
\hat{d}_n=\left ( \frac{\Omega_nB_n}{2\hbar}\right )^\oh
\left ( \alpha_n+\frac{i}{\Omega_n}\dot{\alpha}_n\right )
$$
The collective Hamiltonian will then read
\beq
\hat{H}=\sum_n \hbar\Omega_n\left ( \hat{d}_n^\dagger\hat{d}_n+\oh\right )
\eeq

\section{Electromagnetic structure of ISGDR}

\subsection{Low-$q$ limit of the form-factors multipole parametrization}

To disclose the structure of the ISGDR we adopt the multipolar parametrization 
of charges and currents according to \cite{dubche74}.
In this approach the classical electromagnetic multipoles are expanded in the momentum 
transfer in reactions with photons, electrons or charged hadrons. Let us take 
first the charge multipole form-factor for the charge part of the 
ISGDR fluctuation density:
\beq
M_{\lambda\mu}^{\rm C}(q,t)=\int d\bd{r} j_\lambda(qr)Y_{\lambda\mu}(\vartheta\varphi)
\delta\rho_p(\bd{r},t)\approx \frac{q^{\lambda}}{(2\lambda+1)}
\left( Q_{\lambda\mu}(t)-\frac{1}{2(2\lambda+3)}q^2
\overline{\varrho_{\lambda\mu}^2}(t)\right) 
\label{ffcoul}
\eeq
The first terms of the above $q$-expansion  represents the transition charge dipole
moment
\beq
Q_{\lambda\mu}(t)=\delta_{\lambda,1}\delta_{\mu,0}
\int d\bd{r}r^{\lambda} Y_{\lambda\mu}(\theta,\phi)
\delta\rho_p(\bd{r},t)=0,
\label{diptranmom}
\eeq  
a quantity which vanishes for the ISGDR due to the constraint imposed on the 
c.m. motion. For the IVGDR this will not be the case, since the dynamical dipole moment arises naturally as a measure of the relative motion between the ''negative'' charge (neutron) distribution and the positive charge (proton) distribution. 
Instead the next term in the expansion (\ref{ffcoul}), does not 
cancel. The quantitiy
\beq
\overline{\varrho_{\lambda\mu}^2}(t)=\delta_{\lambda,1}\delta_{\mu,0}
\int d\bd{r}r^{\lambda+2} Y_{\lambda\mu}(\theta,\phi)
\delta\rho_p(\bd{r},t)=2\rho_{0p}R_0^5\sum_n\alpha_n(t)
\frac{j_\lambda(k_L^{(n)}R_0)}{k_L^{(n)}R_0}
\label{squarad}
\eeq
represents the mean square radius of the dipole charge distribution. It provides 
informations on the spatial extension of the ISGDR and it depends only on the longitudinal(compressional) part of the displacement field which are related via the continuity equation (\ref{continuity}) to the density fluctuations.

According to the charge-current multipole parametrization of \cite{dubche74}, 
the electric transverse form factor, splits into the ${q}=0$ limit and a term containing the higher order content in $q$.
\beq
T_{\lambda\mu}^{\rm E}(q,t)=\frac{i^{\lambda+1}}{(2\lambda+1)!!}q^{\lambda-1}
\sqrt{\frac{\lambda+1}{\lambda}}\left( \dot{Q}_{\lambda\mu}(0,t)+
q^2T_{\lambda\mu}^{\rm tor}(q,t)\right) 
\label{ffeltrans}
\eeq  
The first term in the paranthesis is the time-derivative of the Coulomb 
multipole moment defined in eq.(\ref{diptranmom})  
The second term represents the toroidal form factor that reads
in the low-$q$ limit 
\beqa
T_{\lambda\mu}^{\rm tor}(0,t)&=&-\frac{1}{2}\sqrt{\frac{\lambda}{2\lambda+1}}
\int d\bd{r}~r^{\lambda+1}\left[ \bd{Y}_{\lam\lam-1}^\mu+\frac{2}{2\lambda+3}
\sqrt{\frac{\lambda}{\lambda+1}}\bd{Y}_{\lam\lam+1}^\mu\right]\cdot\bd{j}(\bd{r},t) \\
&=&\frac{1}{2i}\sqrt{\frac{\lambda}{\lambda+1}}\frac{1}{2\lambda+3}
\int d\bd{r}~r^{\lambda+2} \bd{Y}_{\lam\lam}^\mu\cdot(\nabla\times\bd{j}(\bd{r},t))
\label{tormom}
\eeqa
In classical electrodynamics the transition toroidal multipole moment is
associated to a poloidal flow on the wings of a toroidal solenoid 
(for details see the reviews \cite{dubche74}).  
In the study of electric collective states it can be related to the strength
of the vorticity associated to a nuclear transition. Indeed, following
ref.\cite{ravwam87}, we introduce the transition multipoles of the 
curl of the current density (unconstrained by the charge-current conservation law) 
$${\cal T}_{\lambda\lambda}(r)\equiv 
\langle I_f\parallel(\nabla\times\bd{j}(\bd{r},t))_{\lambda\lambda}\parallel I_i\rangle$$ 
In order to remove the charge-current conservation constraint, the authors of \cite{ravwam87} introduced the pure vorticity transition multipole
\beq
\omega_{\lambda\lambda}={\cal T}_{\lambda\lambda^{\prime}}(r)-\sqrt{\frac{\lambda+1}{\lambda}}\Omega\rho_{\lambda}
\eeq
where $\rho_{\lambda}$ is the charge density multipole and $\Omega$ the energy 
associated to the transition.
If the quantity
$$
\nu_\lambda=\int_0^\infty dr~r^{\lambda+4}\omega_{\lambda\lambda}(r)
$$
is defined according to \cite{ravwam87} as the strength of the vorticity
and employing the definitions (\ref{squarad}) for the square of the dynamic 
dipole charge distribution  and  (\ref{tormom}) for the dynamic dipole
toroidal moment we arrive at the expression relating the r.m.e. of 
these last two electromagnetic multipoles and the vorticity strength
\beq
\langle I_f=1_n^-\parallel T_{1}^{\rm tor} \parallel I_i=0\rangle
=\frac{i}{10}\left [ \frac{1}{\sqrt{2}}\nu_1
+\Omega_n\langle I_f=1_n^-\parallel \overline{\varrho_{1}^2} \parallel I_i=0\rangle\right]  
\eeq 
From this last formula  we see that since the toroidal dipole moment and the 
square radius of the dipole charge distribution are the leading terms in the $q$-expansion of the transverse electric (\ref{ffeltrans}) and Coulomb (\ref{ffcoul})
form factors for the ISGDR, the determination of these two electromagnetic multipoles 
at low $q$ allows the determination of the vorticity content unconstrained by the
charge-current conservation law.
Before ending this section we give the classical expression of the transition
toroidal dipole moment associated to the ISGDR. Since the current density reads
in this case
\beq
\bd{j}=e\frac{Z}{A}n_0\sum_{n=1}\left( \delta\bd{s}_L^{(n)}(\bd{r})+
r_n\delta\bd{s}_T^{(n)}(\bd{r})\right) 
\eeq
we finally obtain
\beq
T_1^{\rm tor}(0,t)={\frac{1}{10\sqrt{2}}}\rho_{0p}R_0^5\sum_n r_n\dot\alpha_{n}(t)
\frac{j_3(k_T^{(n)}R_0)}{k_T^{(n)}R_0}
\eeq
Thus, the dipole toroidal moment depends only on the shear(vortical) part of the proton
fluid displacement field.

\subsection{Electro-excitation form factors of ISGDR}

Inelastic electron scattering is an excellent tool to explore the nature
of currents involved in the excitation of low-lying (rotational or vibrational) 
and high-lying (giant resonances) collective states  \cite{ueberall}.
A specific feature of this reaction is represented by the possibility
to separate the longitudinal  from the transverse response functions.
This is of vital importance if one tries to disentangle the compresional from the
vortical response in the excitation of a specific electric collective
state.  

\subsection{({\it e,e$^\prime$}) form factors}

Quantizing the expression of the fluctuation density (\ref{denscorr})
and inserting it in the square of the reduced matrix element(r.m.e.) of 
the Coulomb multipole operator (\ref{ffcoul}) leads to
\beq  
\mid \langle I_f=1^-\parallel {\hat M}_{\lambda}^{\rm C} \parallel I_i=0\rangle\vert^2=
3\left[ \rho_{0p}R_0^2\sum_n F_{\rm C}^{(n)}(q)\right]^2  
\label{rmec}
\eeq
where
\beq
F_C^{(n)}(q)=\alpha_{n0}
\left\lbrace \frac{k_l^{(n)}}{q^2-{k_l^{(n)}}^2}
\left[ k_l^{(n)}j_{0}(k_l^{n}R_0)j_{1}(qR_0)
-qj_{0}(qR_0)j_{1}(k_l^{(n)}R_0)\right] -j_{2}(k_l^{(n)}R_0)j_{1}(qR_0)
\right\rbrace \\
\label{ffactcoul}
\eeq
and the amplitude of the canonical coordinate $\alpha_n$ is given by  
the transition r.m.e. of the normal coordinate
\beq
\alpha_{n0}\equiv \langle I_f=1^-\parallel {\alpha}_{n} \parallel I_i=0\rangle
=\left(  \frac{\hbar}{2B_n\Omega_n} \right) ^{1/2}
\eeq
It can be noticed from the content of (\ref{ffactcoul}) that not
only its low-$q$ limit depend on the longitudinal components of the collective flow ($\delta\bd{s}_L$), as substantietad by eq.(\ref{squarad}), but also its higher order
$q$ content. Therefore, the Coulomb form-factor provides solely informations on the compresional waves associated to the ISGDR. 

\begin{figure}
\centerline{\epsfig{file=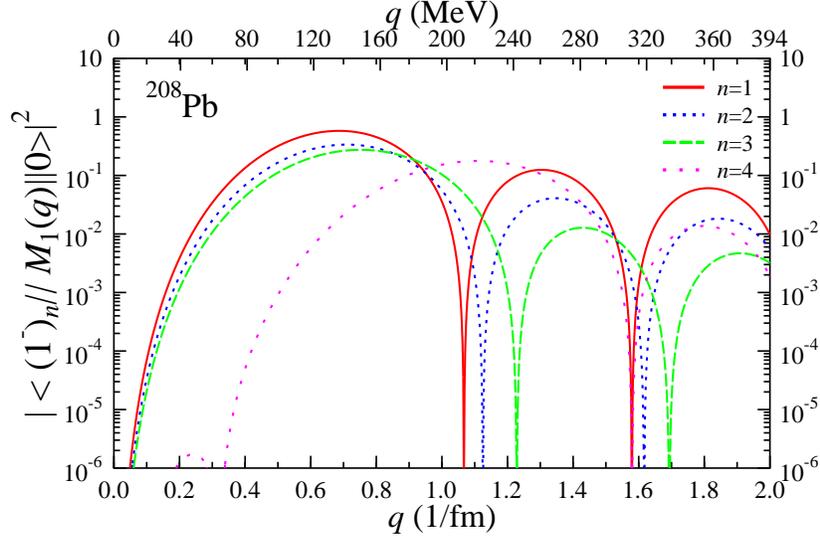,width=11.cm}}
\caption{R.m.e. of the Coulomb multipole operator for the first 4 overtones of the 
ISGDR in $^{208}$Pb.}
\label{fig_rmec}
\end{figure}

In fig.\ref{fig_rmec} we draw the dependence of (\ref{rmec}) on the momentum
transfer $q$ for the first four overtones in $^{208}$Pb.  The diffraction maxima, although shifted to higher values $q$ for increasing overtone number, are not affected in absolute value.

\begin{figure}
\centerline{\epsfig{file=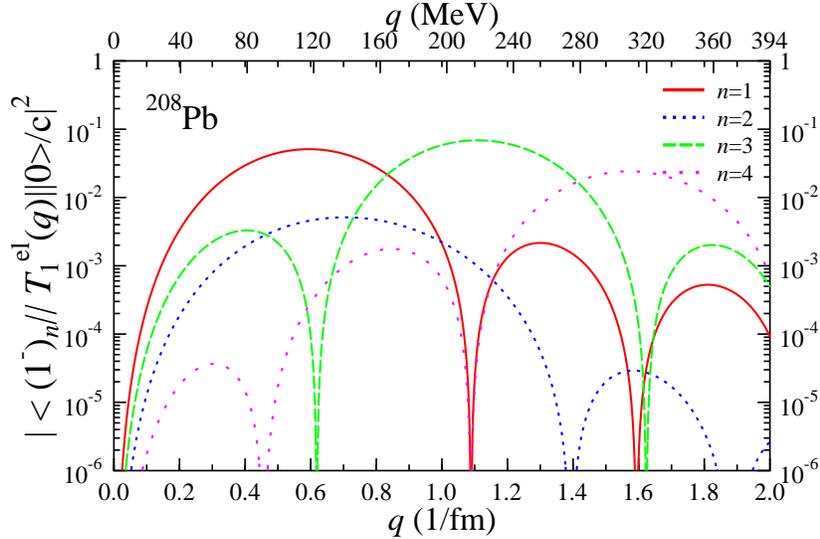,width=11.cm}}
\caption{R.m.e. of the transverse multipole operator for the first 4 overtones of the 
ISGDR in $^{208}$Pb.}
\label{figrmtel}
\end{figure}

The expression of the squared r.m.e. of the transverse electric operator (\ref{ffeltrans}) reduces to
\beq
\mid \langle I_f=1^-\parallel \widehat{T}^{\rm E}_{\lambda}(q) \parallel I_i=0\rangle\vert^2 = 3\left[\rho_{0p}R_0^3\sum_n 
F_{\rm T}^{(n)}(q)\right ]^2
\label{transvop}
\eeq
where
\beq
F_T^{(n)}(q)=r_n\Omega_n\alpha_{n0}\left\lbrace 
R_0\frac{k_T^{(n)}j_{1}(k_T^{n}R_0)j_{0}(qR_0)
-qj_{1}(qR_0)j_{0}(k_T^{(n)}R_0)}{q^2-{k_T^{(n)}}^2}
-\frac{3}{qk_T^{(n)}}j_{1}(k_T^{(n)}R_0)j_{1}(qR_0)
\right\rbrace
\eeq
Thus, contrary to the Coulomb multipoles, the transverse multipoles
are providing informations only on the vortical components of the ISGDR.
The squared r.m.e. of (\ref{transvop}) is ploted in Fig.\ref{figrmtel}.
The first diffraction bump dominates for the $n=$1 and 2 components of the ISGDR, 
whereas for $n$=3 the second bump and for $n=4$ the third bump are taking over. 
According to Table 
\ref{tabela1}, were we listed the ratio $b_n/a_n$(vorticity/compressibility),
this feature, is a consequence of the vorticity enhancement,  in this higher lying 
states.  
Instead the first Coulomb diffraction bump maintains its importance in the electro-excitation of higher-lying states because the role played by the density waves is less important for these states.
 
\subsection{Sum rules for electron scattering}

It is known that for the isoscalar electric modes simulated by operators depending only on coordinates of the particles the energy-weighted sum rules can be determined model independently and depend only on the ground-state properties of the nucleus
\cite{boh75,deal73,har81,lipstr89}. 
Operators from this class are commuting with interactions not depending explicitely 
on the momenta of the particles.  These scalar operators are simulating shape or density distortions corresponding to a given isoscalar multipolar resonance and for that 
reason in the literature the macroscopic images associated to these excitations are always irrotational surface or bulk compressional oscillations.
Instead they are inadequate to describe distortions of the nuclear current which are not constrained by the charge-current conservation law, i.e. excitations with vortical 
currents.  A class of sum-rules coping with both kind of distortions, i.e.
of charge density and current (unconstrained by the charge-current relations) is given
by the longitudinal ($L$) and transverse ($T$) intrinsic energy weighted sum-rules (EWSR) at constant three-momentum transfer $|\bd{q}|$, which are constructed by weighting the nuclear response functions \cite{torn80} 
\beqa
R^L(\bd{q},E^*)&=&\sum_n^\infty\mid\langle n \mid \rho(\bd{q})\mid 0\rangle\mid^2
\delta\left( E^*-\frac{\hbar^2q^2}{2M_A}-E_n\right) \\
R^T(\bd{q},E^*)&=&\sum_n^\infty\mid\langle n \mid \bd{J}_\bot(\bd{q})\mid 0\rangle\mid^2
\delta\left( E^*-\frac{\hbar^2q^2}{2M_A}-E_n\right)
\eeqa 
with an appropriate power of the nuclear excitation energy $E^*$. Summing over 
all excited states we obtain the intrinsic $p$-order EWSR depending on $q$
\beqa
{m_p^L}(q)&=&\int d{E^*}' R^L(\bd{q},{E^*}'){E^*}^p= \sum_n^\infty E_n^p\langle n \mid \rho(\bd{q})\mid 0\rangle\mid^2
\label{sumrulong}\\
{m_p^T}(q) &=&\int dE^{*\prime}{R^T}(\bd{q},E^{*\prime}) {{E}^{*\prime}}^p= 
\sum_n^\infty {E_n^p}\langle n \mid \bd{J}_\bot(\bd{q})\mid 0\rangle\mid^2
\label{sumrutran}
\eeqa
where $E^{*\prime}={E}^*-\hbar^2q^2/2M_A$ is the energy available for intrinsic excitations. In the above formula $\bd{J}_\bot(\bd{q})$ 
denotes the transverse component of the current operator relative to the momentum transfer$(\bd{J}_\bot(\bd{q})=\bd{J}(\bd{q})-\bd{q}(\bd{q}\cdot\bd{J})/\bd{q}^2)$.

\begin{figure}[t]
\centering
\mbox{\subfigure{\epsfig{file=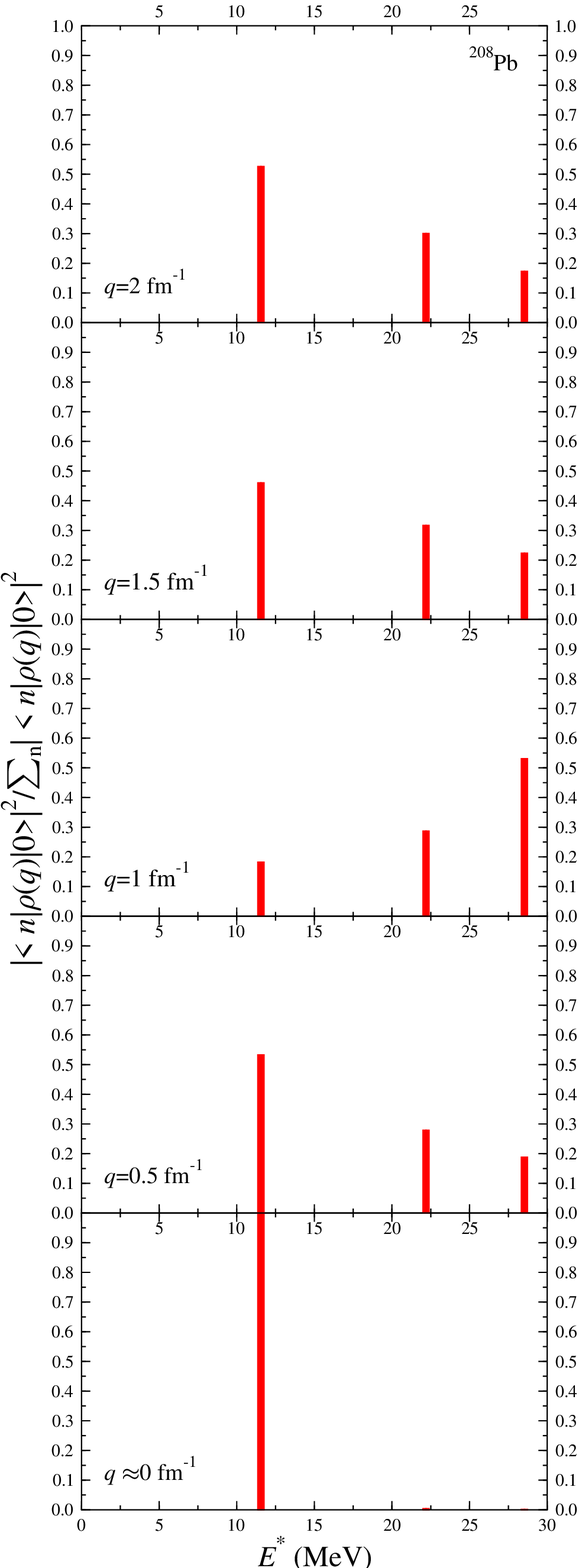,width=7.7cm}}
       \subfigure{\epsfig{file=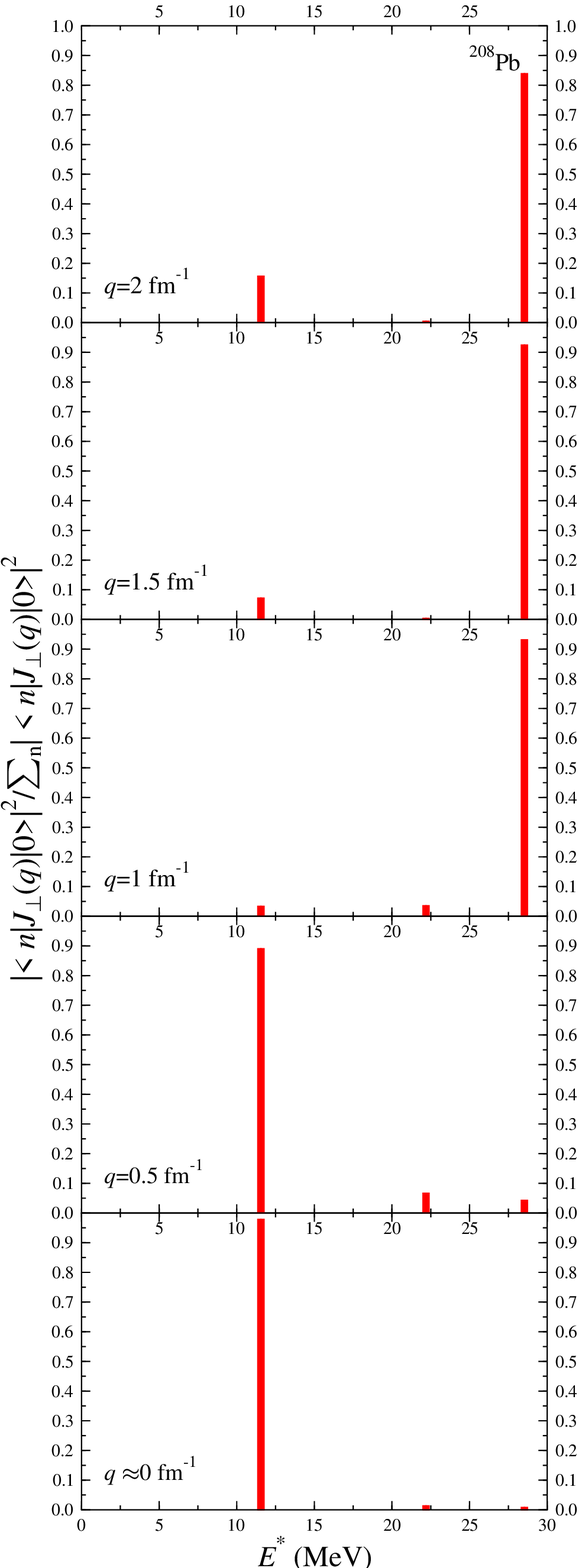,width=7.7cm}}
}
\caption{Zeroth-order energy strengths distributions for $\rho$
and $\bd{J}_\perp$. The energy cut is $(E^*)_{\rm cut}$ = 30 MeV.}
\label{ewsr0}
\end{figure}

\begin{figure}[t]
\centering
\mbox{\subfigure{\epsfig{file=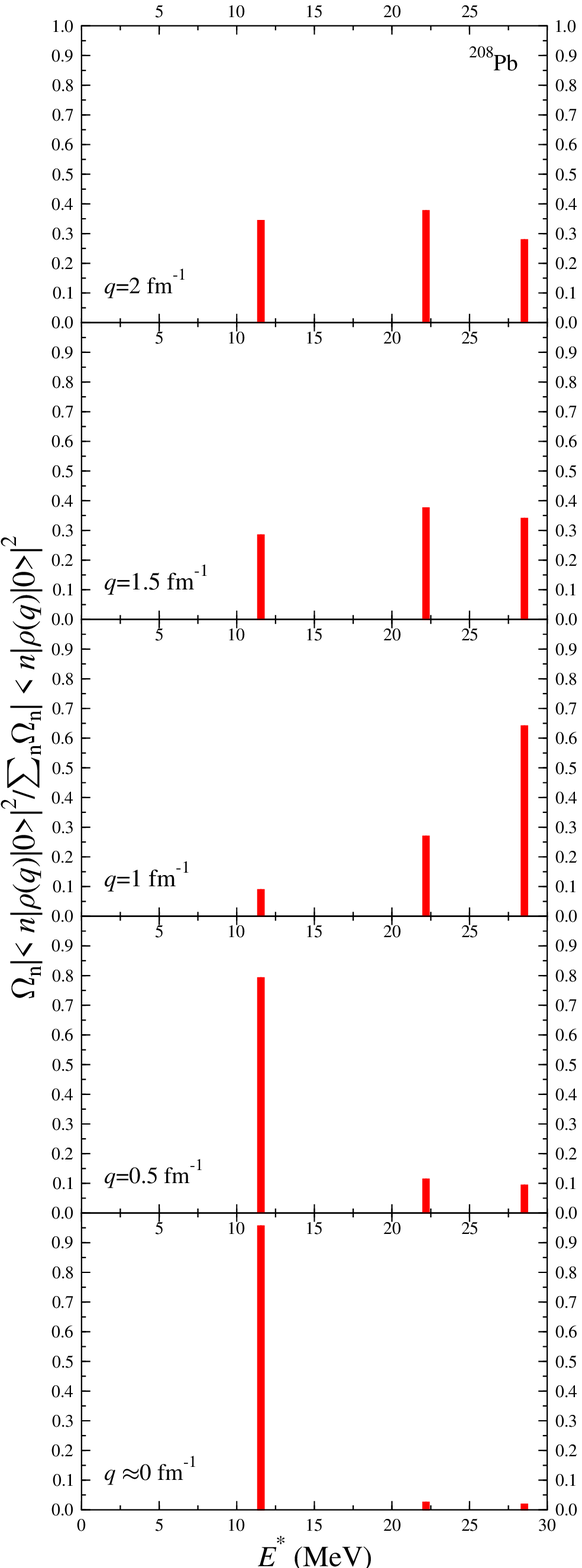,width=7.7cm}}
       \subfigure{\epsfig{file=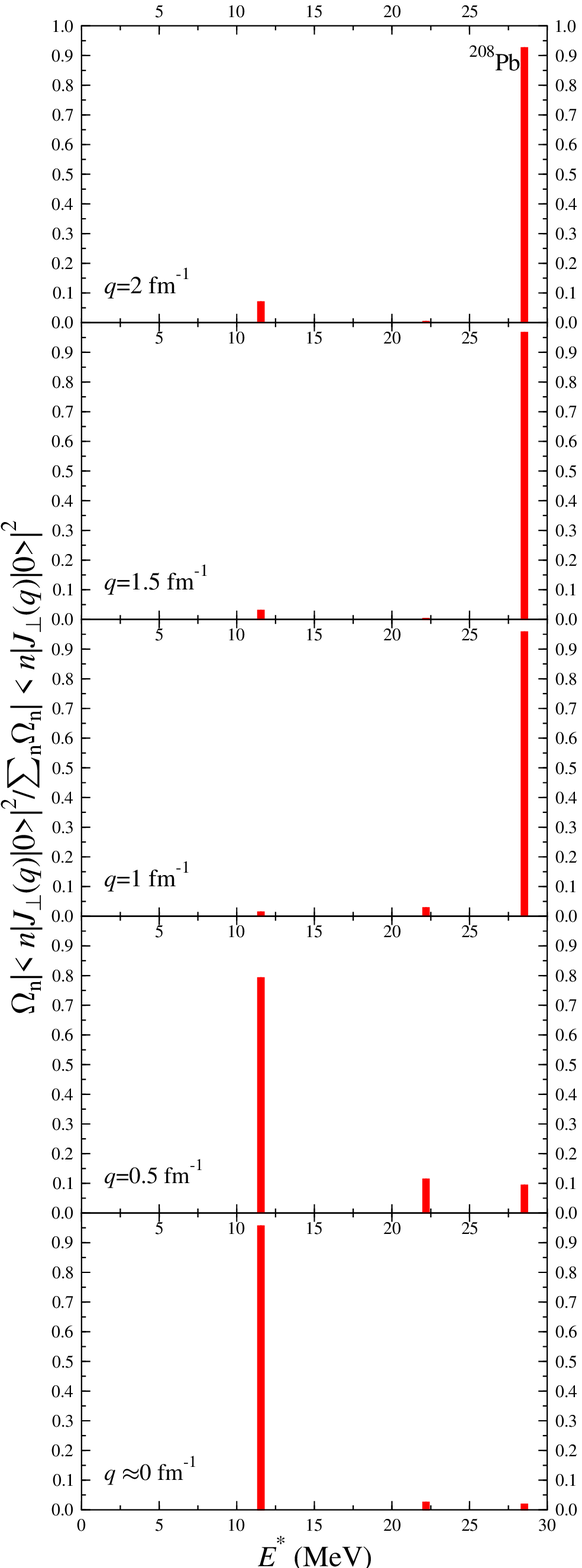,width=7.7cm}}
}
\caption{First-order energy strengths distributions for $\rho$ and
$\bd{J}_\perp$. The energy cut is $(E^*)_{\rm cut}$ = 30 MeV.}
\label{ewsr1}
\end{figure}

The longitudinal and transverse $p-$order energy strengths of each state can then be obtained as relative contributions to the corresponding sum-rules (\ref{sumrulong}) 
and (\ref{sumrutran})
$$E_n^p\mid\langle n \mid \rho(\bd{q})\mid 0\rangle\mid^2/{{m_p}^L}(q),~~~~~
E_n^p\mid\langle n \mid \bd{J}_\bot(\bd{q})\mid 0\rangle\mid^2/{{m_p}^T}(q)$$
In Figs.\ref{ewsr0} and \ref{ewsr1} we represent the zeroth- and first- order strengths distributions for the longitudinal and transverse responses.  In the sums only the first three overtones were included and therefore the excitation energy is 
truncated at 30 MeV. We see that for very low momentum transfer the $L$ and $T$ strength
is mainly concentrated on the first overtone.   When $q$ increases the $L$ and $T$
strengths follow a different pattern. Whilst the $L$ strength is fragmented almost ''democratically'' over the three overtones, the $T$ strengths are testifying a transition
from a low-$q$ regime where the first overtone dominates to a high-$q$ regime where
the third overtone overtakes the predominance. For both regimes the second overtone
plays a very minor role. This fact can be explained in our view by the compressibility
content of this mode which is "washed-out" in the transverse response function.
In the longitudinal response  the second overtone plays a more visible role.

In the low-$q$ limit the ratios $m_1/m_0$ and $\sqrt{m_3/m_1}$ are expected to provide crude estimates for the mean excitation energies associated to the density or transverse density current distortions. In Table \ref{tabela2} we list these mean excitation energies for a series of spherical nuclei. Comparing the obtained values to the 
eigenvalues of the first two overtones we notice that the ratio 
${m_1}^L/{m_0}^L$ gives the best estimate of the first overtone.  For the  
second overtone the ratio $\sqrt{{m_3}^T/{m_1}^T}$ gives a rather coarse 
approximate. The other two ratios are falling in the vicinity of the first overtone.

\begin{table}
\begin{center}
\begin{tabular}{ccccccccc}
\hline
\hline
Nucleus& ${m_1}^L/{m_0}^L$ & $\sqrt{{m_3}^L/{m_1}^L}$ & ${m_1}^T/{m_0}^T$
& $\sqrt{{m_3}^T/{m_1}^T}$ & $\hbar\Omega_{1}$&$\hbar\Omega_{2}$
& $(\hbar\Omega_{1})_{\rm exp}$&$(\hbar\Omega_{2})_{\rm exp}$\\
&(MeV)&(MeV)&(MeV)&(MeV)&(MeV)&(MeV)&(MeV)&(MeV)\\
\hline
$^{90}$Zr&15.44& 16.29& 16.86& 25.53& 15.28 & 29.34&16.20$\pm$0.80&25.70$\pm$0.70$^{1)}$\\
$^{116}$Sn& 14.19& 14.99& 15.51& 23.68& 14.04 & 26.96&14.38$\pm$0.25&25.50$\pm$0.60$^{1)}$\\
                                              &&&&&&&14.70$\pm$0.80&23.00$\pm$0.60$^{2)}$\\
$^{144}$Sm& 13.21& 14.07& 14.64& 23.95& 13.07 & 25.08&14.00$\pm$0.30&24.51$\pm$0.40$^{1)}$\\
$^{208}$Pb& 11.68& 12.44& 12.96& 21.19& 11.56 & 22.19&13.26$\pm$0.30&22.20$\pm$0.30$^{1)}$\\
 &&&&&&&12.50$\pm$0.30&22.50$\pm$0.30$^{2)}$\\
\hline
\hline
\end{tabular}
\caption{Ratios of ISGDR sum rules for $q\longrightarrow 0$ compared
to the energies of the first two overtones for $^{90}$Zr, $^{116}$Sn, $^{144}$Sm
and  $^{208}$Pb and with experimental latest experimental data  from \cite{young04}$^{1)}$ and
\cite{uch03}$^{2)}$ for the low- and high-energy peaks .}
\label{tabela2}
\end{center}
\end{table}

\section{Summary and Conclusion}

The lower and upper component of the ISGDR, as reported by the latest
experimental measurements, are explained 
as the first two overtones of a spherical Fermi-Fluid system with a sharp free surface
corresponding to a mixture of dipole compression and vorticity oscillations.

The approach presented in this work was applied primary to the heavy 
spherical nucleus $^{208}$Pb, because in this case the sharp-edge density
distribution assumption is more acceptable as would be the case for lighter nuclei
for which experimental data on ISGDR are available ($^{40}$Ca, $^{90}$Zr, 
$^{116}$Sn and $^{144}$Sm) and where the diffuse surface plays an important
role. In order to extend the analysis of ISGDR to these  nuclei and also to 
exotic nuclei that are presently under intense investigations, one should adopt 
at first a more realistic assumption for the ground-state density distribution.  
In this case along with the density, other parameters of the nuclear Fermi liquid.
e.g. the Lam\' e coefficients acquire a radial dependence and
the equations of motions must be solved numerically.

In the present approach, the excitation of the ISGDR is not limited to the 
squeezing operator. It includes the entire momentum content in the 
operator $j_1(kr)Y_1({\hat r})$ and takes into account also its c.m. correction
via constraints on the density and displacement field fluctuations. 
Moreover, since there is no mathematical or physical exception it considers along with
the longitudinal solution also the transverse solution of the vector Helmholz equation.
Conseqently the overtones of the ISGDR are mixtures of  compressional and vortical velocity fields. For the second overtone, previously advocated to be of compressional nature, the nuclear Fermi-Fluid approach confirms very recent microscopical predictions \cite{cro-ring01,kvasil03} that are pointing toward a cohabitation of compressional and 
vorticity vibrations in the ISGDR states up to 30 MeV. 

Naturally, a question arise : Has been gained so far any indication in experiment 
on the overtones with $n\geq 3$, located above the ''high-lying'' ISGDR state?
For the time being this question cannot be answered because the latest data reported by the College Station group \cite{young04} are providing very large uncertainties in the strength distribution for $^{208}$Pb beyond the second peak, i.e. at excitation energy 
$>3\hbar\omega$.  

A study of the electromagnetic multipole transitions was undertaken and concluded that the leading term in the Coulomb form-factor is singled-out only by the density fluctuation 
and can be related to the rms-charge radii. In turn the leading mutipole in the transverse
electric form factor, the toroidal dipole moment, results soleley from the 
transverse part of the velocity field. In this respect the r.m.e. of the 
toroidal dipole transition provide a signature of vorticity through electromagnetic
probes.

The $(e,e^\prime)$  are promising candidates for the exploration of the role
of longitudinal and vortical currents of ISGDR since the separation of longitudinal
(Coulomb) and electric transverse form-factor is feasible. However in such reactions
it is difficult to avoid the excitation of the dominant electric dipole response, the 
IVGDR. It would be in this respect interesting to search for a macroscopical description
that deals in a unified manner with the isoscalar and isovector electric dipole responses.

\begin{acknowledgments}
The author(\c S.Mi\c sicu) aknowledge the 
financial support from the Alexander von Humboldt Foundation and the hospitality 
of Prof.A.Richter at the Institut f\"ur Kernphysik, TU-Darmstdat, where this work 
was accomplished. The author is also very gratefull to Prof.P.v.Neumann-Cosel and
Dr.V.I.Ponomarev for usefull discussions.
\end{acknowledgments}

\appendix
\section{Scalar and vector Helmholz equations}

The solution of the scalar HE (\ref{helmhol}) reads
\beq
{\cal D}=\sum_{\lam\mu}a_{\lam\mu}j_\lam(k_Lr)Y_{\lam\mu}(\theta,\phi)
\eeq
The solution of the vector HE splits into an electric(poloidal) and a 
magnetic(torsional) solution
\beqa
\bd{\omega}^{\rm pol}&=&\sum_{\lam\mu}b_{\lam\mu}^{\rm el}j_\lam(k_Tr)
\bd{Y}_{\lambda\lambda}^{\mu}(\theta,\phi)\\
\bd{\omega}^{\rm tor}&=&\sum_{\lam\mu}b_{\lam\mu}^{\rm el}
\frac{1}{\sqrt{2\lambda +1}}
\left ( \delta_{\lambda'\lambda -1}\sqrt{\lambda +1} -
        \delta_{\lambda'\lambda +1}\sqrt{\lambda}\right )
j_{\lambda'}(k_T r)\bd{Y}_{\lambda\lambda'}^{\mu}(\theta,\phi)
\eeqa
In the present study on electric resonances we are interested only 
in the poloidal solution. 
Using the properties of the spherical harmonic vectors we can derive the 
expression of the longitudinal and transverse displacements fields.
Since $\delta\bd{s}_L$ results from the definition of the scalar function
\beq
{\cal D}\equiv\nabla\cdot\delta\bd{s} =\nabla\cdot\delta\bd{s}_L
\eeq
we have that
\beq
\delta\bd{s}_L=-\frac{1}{k_L}\sum_{\lam\mu}a_{\lam\mu}
\frac{1}{\sqrt{2\lambda +1}}
\left ( \delta_{\lambda'\lambda -1}\sqrt{\lambda' +1} +
        \delta_{\lambda'\lambda +1}\sqrt{\lambda'}\right )
j_{\lambda'}(k_L r)\bd{Y}_{\lambda\lambda'}^{\mu}(\theta,\phi)
\label{app-long}
\eeq
Similarly, from the definition of the vorticity
\beq
\bd{\omega}^{\rm pol}\equiv\oh\nabla\times\delta\bd{s}_T
\eeq
we get
\beq
\delta\bd{s}_T=\frac{i}{k_T}\sum_{\lam\mu}b_{\lam\mu}
\frac{1}{\sqrt{2\lambda +1}}
\left ( \delta_{\lambda'\lambda +1}\sqrt{\lambda'} -
        \delta_{\lambda'\lambda +1}\sqrt{\lambda'-1}\right )
j_{\lambda'}(k_T r)\bd{Y}_{\lambda\lambda'}^{\mu}(\theta,\phi)
\label{app-tran}
\eeq

\end{document}